\documentclass{aastex}
\usepackage{spr-astr-addons}

\usepackage{url,graphicx}

\RequirePackage{color}

\begin{document}

\newcommand{\zav}[1]{\left(#1\right)}

\title{NLTE wind models of hot subdwarf stars}
\shorttitle{NLTE wind models of hot subdwarf stars}
\shortauthors{Krti\v cka \& Kub\'at}

\author{Ji\v r\'\i\ Krti\v cka\altaffilmark{1}}
\affil{\'Ustav teoretick\'e fyziky a astrofyziky, Masarykova univerzita, CZ-611 37 Brno, Czech Republic}
\and
\author{Ji\v r\'\i\ Kub\'at\altaffilmark{2}}
\affil{Astronomick\'y \'ustav, Akademie v\v{e}d \v{C}esk\'e republiky, CZ-251 65 Ond\v{r}ejov, Czech Republic}

\begin{abstract}
We calculate NLTE models of 
%
stellar winds of hot compact stars
(central stars of planetary nebulae and subdwarf stars). 
The studied
range of
subdwarf parameters is selected to cover a large part of these
stars. 
The models
predict
the
wind hydrodynamical structure and provide mass-loss 
rates
for different abundances. 
Our models show that
CNO elements are important
drivers of subdwarf winds, especially for low-luminosity stars. We study the
effect of X-rays and instabilities on these winds. Due to the line-driven wind
instability, a significant part of the wind could be very hot.
\end{abstract}

\keywords{stars: winds, outflows; stars: mass-loss; stars:
    early-type;  stars: horizontal-branch; planetary nebulae: general}


\section{Subdwarf stars and the stellar wind}

Despite 
being
low-luminous, the horizontal branch stars may still have stellar
winds
due to their large effective temperatures and low stellar 
radii.
%
The winds
of these stars is thus similar to 
those
of luminous OB stars
\citep{vinca,unglaub}, which are driven mainly 
by photon
absorption in lines
of 
heavy
elements, e.g., C, N, O, or Fe 
(see \citealt{owopo}; \citealt{kkpreh}; \citealt{pulvina} for a review; or
\citealt{wrh}).

The main wind parameters are 
the
mass-loss rate $\dot M$ and 
the
terminal velocity
$v_\infty$. The 
%
mass-loss rate depends on 
the
stellar luminosity $L$
as
$\dot M
\sim L^{1/\alpha'}$ \citep{pulvina}, where $\alpha'\approx0.6$ for O stars
\citep{nlteiii}. This scaling implies that the 
%
mass-loss rate increases
with luminosity. Because the wind is driven by metal
%
lines, the 
%
mass-loss rate depends also on 
the
mass fraction of heavier elements, $Z$. For O
stars the scaling $\dot M \sim Z^{0.67}$ is roughly valid \citep{nlteii}. 
In
contrast, the wind terminal velocity is 
proportional to the
surface escape speed,
$v_\infty\propto v_\text{esc}$.

Here we complement the studies of hot subdwarf winds provided by \citet{vinca}
and \citet{unglaub}.

\section{Wind models}

The stellar
radiative flux 
(at the
inner wind boundary) is taken from static,
spherically symmetric NLTE model atmo\-sphe\-res calculated 
with
the code of
\citet{ATAmod}.

For calculation of wind models we apply the NLTE wind code of \citet{nltei}. The
wind is modelled assuming
a
spherically symmetric, stationary flow.
%
Ionization and excitation 
are
derived from the solution of
the
statistical
equilibrium equations. Ionic models adopted from the TLUSTY code
\citep{bstar2006} are based on data from the Opacity and Iron Projects
\citep{topt,zel0}.

The derived excitation and ionization 
are
used to calculate 
the
radiative force
and radiative heating.
%
These terms are inserted in
the
corresponding
hydrodynamical equations (continuity, momentum, and energy). These equations may
be solved for each component of the flow separately. 

From 
the
obtained wind structure we calculate 
the
mass-loss rate and 
the
terminal
velocity. These values can be compared with parameters derived from
observations.

\section{Stellar 
winds
of central stars of planetary nebulae}

\begin{figure}[tp]
\includegraphics[width=0.48\textwidth]{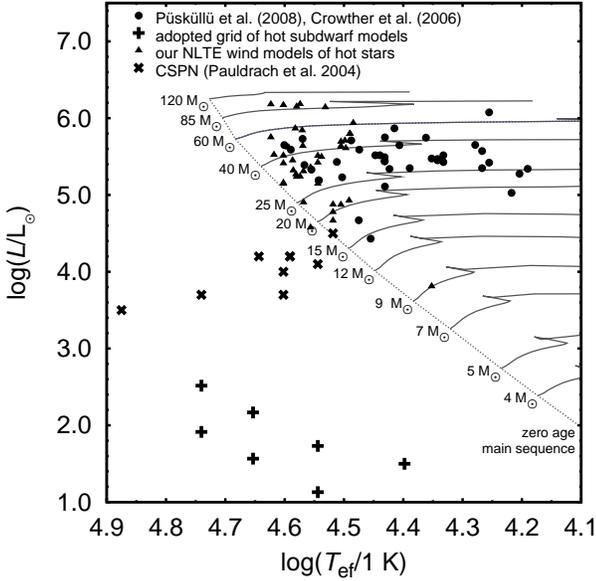}
\caption{%
The position of 
the
studied stars in 
the
HR diagram, compared to luminous O stars}
\label{hrd}
\end{figure}

Up to now, the wind models of \citet{nltei} were applied only to stellar 
winds
of
luminous O stars. Because there is a large gap between 
the
parameters of luminous O
stars and 
O subdwarfs
(see Fig.~\ref{hrd}), we first decided to calculate
wind models of central stars of planetary nebulae (CSPN). This enables us to
test the applicability of our wind models 
%
for stars with low luminosity.

For this test we 
chose
the dataset of \citet{btpau}, who derived the stellar
and wind parameters of selected CPSNs from observations.

\begin{figure}[tp]
\includegraphics[width=0.45\textwidth]{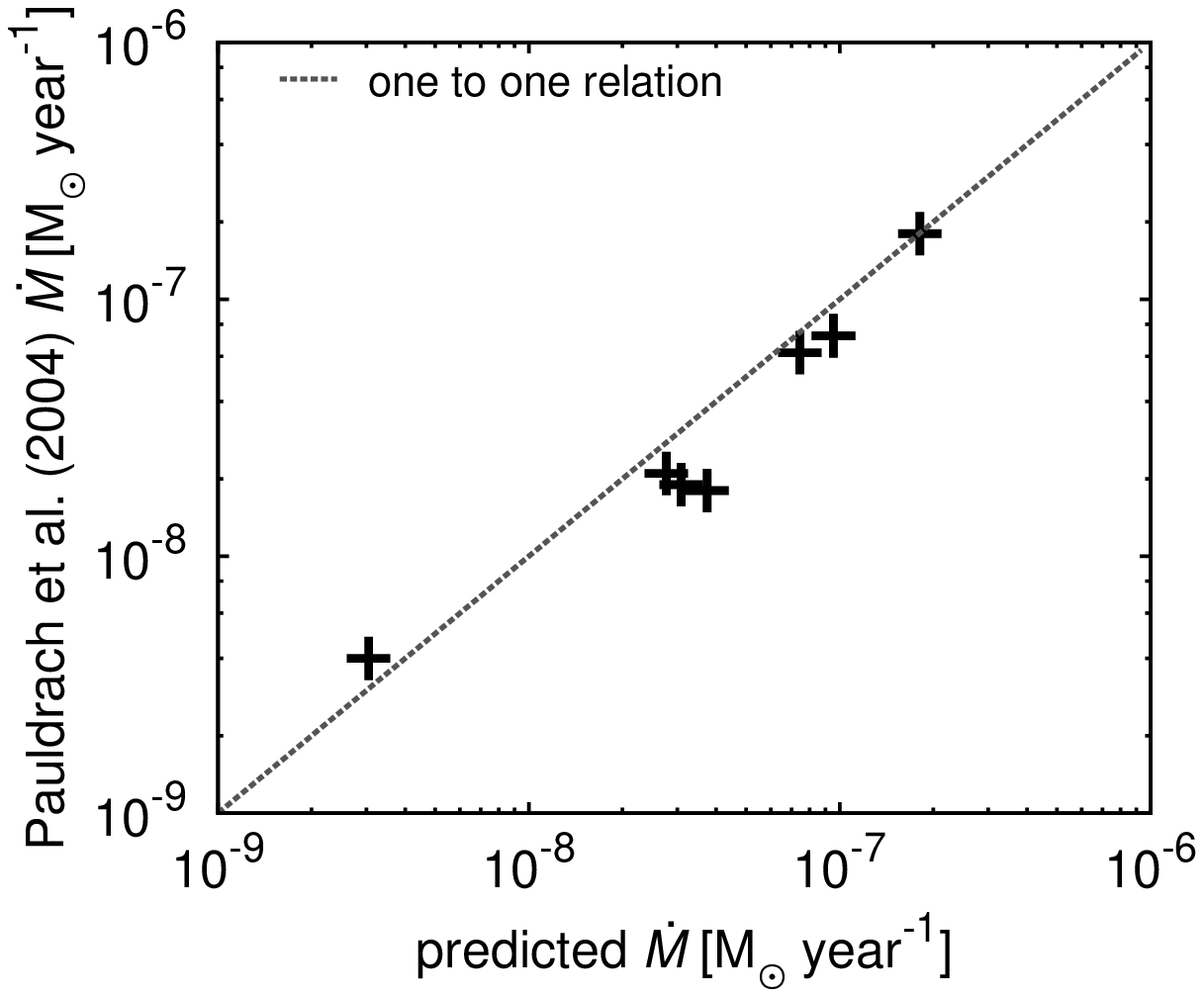}
\includegraphics[width=0.45\textwidth]{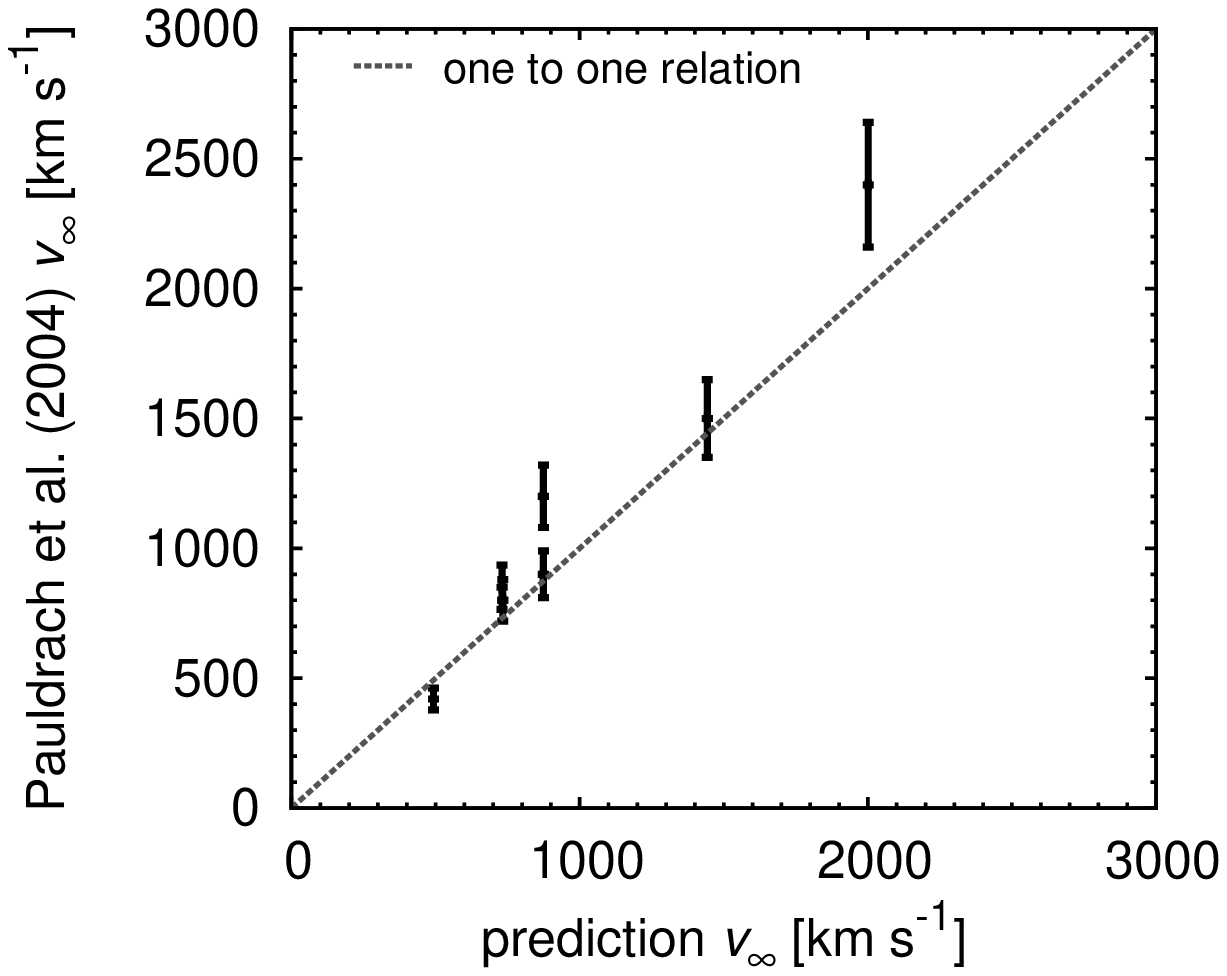}
\caption{%
Comparison of predicted mass-loss rates (upper panel) and terminal velocities
(lower panel) with results derived by \citet{btpau} for 
CSPNe.}
\label{vnekcspn}
\end{figure}

A comparison between
wind mass-loss rates and terminal velocities derived by us and by
\citet[see Fig.~\ref{vnekcspn}]{btpau} 
showed good agreement.
%

\begin{figure}[hp]
\includegraphics[width=0.45\textwidth]{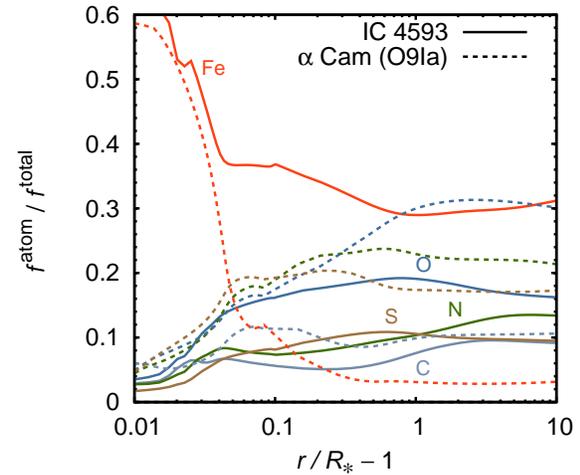}
\caption{%
The radial dependence of 
the
contribution 
from
individual elements to the radiative
force in the wind of CSPN IC~4593 in comparison to the wind of $\alpha$~Cam}
\label{ic4593grelo}
\end{figure}

The
contribution of individual elements to 
the radiative force (Fig.~\ref{ic4593grelo})
shows that iron contributes to 
the
radiative force particularly close to the star.
Lighter elements are important in the outer wind regions. This is similar to
stellar 
winds
of luminous O stars. We conclude that stellar wind of CSPN
resembles that of luminous O stars.

\section{Subdwarf stars}

\begin{figure}[p]
\includegraphics[width=0.45\textwidth]{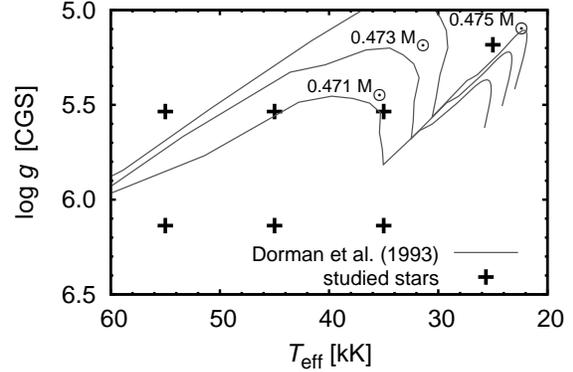}
\caption{%
Stellar parameters of 
our
grid of subdwarf 
models}
\label{sitpot}
\end{figure}

Turning now our attention to subdwarf stars, we selected a grid of effective
temperatures
$T_\text{eff}$ and surface gravity acceleration $g$ that well covers
the area of subdwarf parameters (see Fig.~\ref{sitpot}). We assumed 
a
fixed
stellar mass $M=0.5\,\text{M}_\odot$,
%
different values of metallicity
$0.3\,Z_\odot$, $1\,Z_\odot$, $3\,Z_\odot$, $10\,Z_\odot$, and 
helium-to-hydrogen ratios
$N(\text{He})/N(\text{H})$ 
of
$0.085$, and $100$
(by number).
%

\begin{figure}[p]
\includegraphics[width=0.45\textwidth]{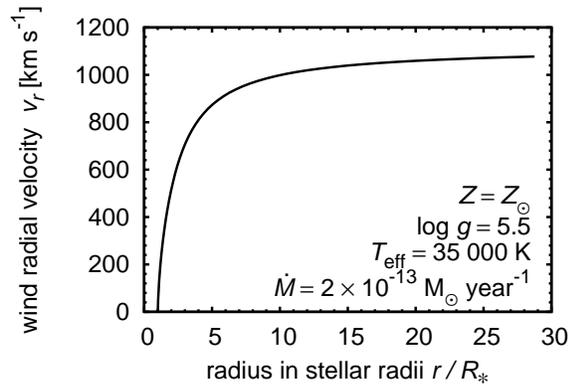}
\caption{%
The dependence of the wind radial velocity on 
%
radius in a selected 
%
model}
\label{3502Z01}
\end{figure}

\begin{figure}[p]
\includegraphics[width=0.45\textwidth]{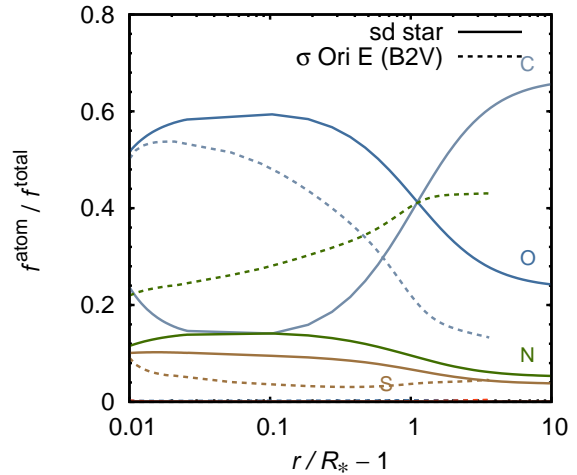}
\caption{%
The radial dependence of 
the
contribution 
from
individual elements to the radiative
force in the winds of 
a
subdwarf star with $T_\text{eff}=35\,000\,\text{K}$, $\log
g=5.5$, $Z=Z_\odot$,
compared to $\sigma$ Ori E}
\label{3502Z01grelb}
\end{figure}

\begin{figure}[pt]
\includegraphics[width=0.45\textwidth]{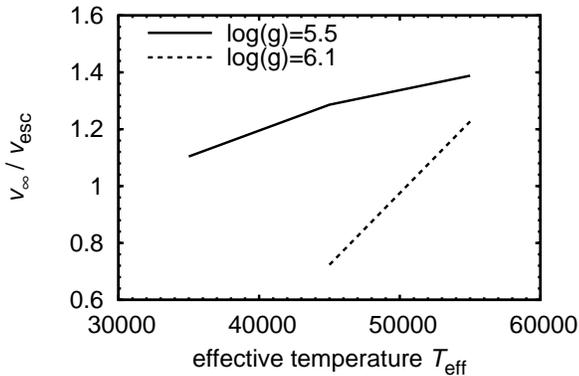}
\caption{%
The ratio of the wind terminal velocity and the surface escape speed for
$Z=Z_\odot$}
\label{tefvnek}
\end{figure}

The
low luminosity of subdwarf stars implies also 
%
low 
%
mass-loss 
rates (see
Fig.~\ref{3502Z01}). Contrary to the wind of 
CSPNe, the
radiative force is
dominated by lines of lighter elements everywhere, similarly to the wind of
early B main-sequence stars (see Fig.~\ref{3502Z01grelb}). 
The
wind terminal
velocity $v_\infty$ 
%
is of the
order of 
the
surface escape speed $v_\text{esc}$ 
(Fig.~\ref{tefvnek}).

\section{Contribution of different elements}

To understand how different elements contribute to 
the
radiative force in different
situations, let us mention that the radiative force due to a particular line is
given by 
the
line optical depth and 
the
radiative flux. 
The
line optical depth in 
%
Sobolev approximation is
\begin{equation}
\tau_\text{S}\sim f_{ij} n_i \zav{\frac{\text{d} v_r}{\text{d}
r}}^{-1},
\end{equation}
where $f_{ij}$ is the oscillator strength, $n_i$ is 
the
number density of atoms in a
given state, and $\frac{\text{d} v_r}{\text{d} r}$ is the velocity gradient
\citep{pulvina}. 
The
radiative force due to optically thick lines
($\tau_\text{S}>1$) does not depend on level populations $n_i$, whereas
the
radiative force in optically thin lines ($\tau_\text{S}<1$) depends on level
populations.

To describe 
the
contribution of individual lines to the radiative force, let us
divide them 
into
strong ones (typically resonance lines of dominant ionization
stages of C, N, O), intermediate ones (strongest iron lines), and weak ones
\citep[cf.][]{vikow}. For high wind density both strong and intermediate lines
are optically thick, consequently each such line contributes by 
a
similar amount
(given by 
the
radiative flux) to the radiative force. However, because there are
much more iron lines than lines of CNO elements, iron lines dominate at large
densities. On the other hand, at low densities strong lines may be still
optically thick (and consequently contribute to the radiative force), whereas
iron lines become optically thin. As the radiative force 
by
optically thin lines
depends on 
the
abundance
of its element,
which is relatively low in the case of iron, 
the
contribution
of iron to the radiative force is small in 
a
low-density environment
\citep{pusle,vikolamet,kkiv}.

The stellar 
winds
of 
CSPNe
or O stars 
have
relatively high density close to the
star and low density in the outer parts.
Consequently,
the iron lines dominate
the radiative driving for small radii, whereas the lighter elements are
important at large radii (see Fig.~\ref{ic4593grelo}). On the other hand, the
stellar wind of subdwarf stars or main sequence B stars has relatively low
density everywhere. 
Consequently,
lines of light elements dominate at all radii
(see Fig.~\ref{3502Z01grelb}).

\section{Multicomponent wind structure}

\begin{figure}[t]
\includegraphics[width=0.45\textwidth]{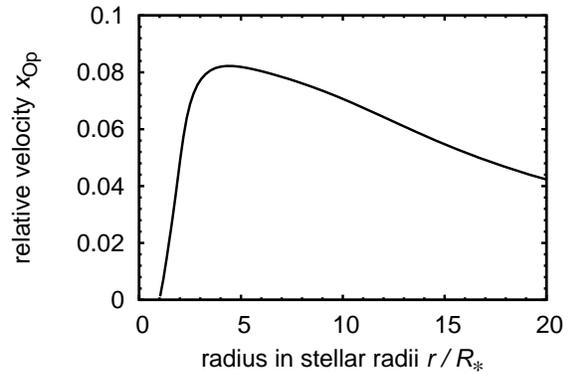}
\includegraphics[width=0.45\textwidth]{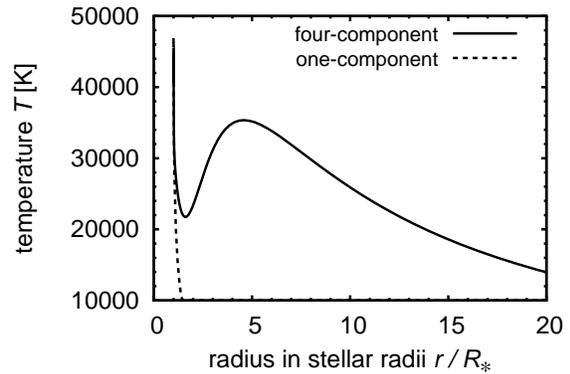}
\caption{%
Frictional heating in models of the star with
$T_\text{eff}=55\,000\,\text{K}$, $\log g = 6.1$, and $Z=3Z_\odot$. Top: the
non-dimensional velocity difference Eq.~\eqref{xip}. Bottom: Comparison of wind
temperatures on one-component and four-component models}
\label{55-01Z034k}
\end{figure}

There are more similarities between the stellar 
winds
of hot subdwarf stars and
%
of early main-sequence B stars. One of them is connected with inefficient
transfer of momentum from 
the
radiatively driven ions to 
the
passive ions of hydrogen
and helium.

In line driven winds, momentum from radiation is transfered to heavier ions,
whereas hydrogen and helium 
practically do not receive 
any momentum from the
radiation field. Because heavier elements constitute only a very small part of
the stellar wind (of about 1\% in mass), momentum shall be transferred from
these heavier ions to 
the
passive ones. As the stellar wind of hot stars is ionized,
Coulomb collisions provide the most 
efficient
way of such transfer of momentum.

For stars with relatively high wind-density,
the
transfer of momentum between
individual heavier ions (denoted by $\text{i}$) and 
the
hydrogen-helium component
($\text{p}$) is efficient, because 
the
velocity difference between these components
is relatively small. However, for low density stellar winds the momentum
transfer becomes inefficient and two effects may emerge: frictional heating and
decoupling of wind components \citep[e.g.,][]{kkii,op,ufo,unglaub}.

The importance of these effects can be assessed from the value of the
non-dimensional velocity difference
\begin{equation}
\label{xip}
x_\text{ip}=\frac{|{v}_\text{i}-{v}_\text{p}|}{\alpha_\text{ip}},
\end{equation}
where ${v}_\text{i}$, and ${v}_\text{p}$ are 
the
velocities of the components,
%
\begin{equation}
\alpha_\text{ip}^2=\frac{2k\zav{m_\text{i}T_\text{p}+
m_\text{p}T_\text{i}}} {m_\text{i}m_\text{p}},
\end{equation}
is the mean thermal speed,
where $m_\text{i}$, and $m_\text{p}$ are
the
ionic masses, and $T_\text{i}$, $T_\text{p}$
the
corresponding temperatures.

For small velocity differences, $x_\text{ip}\lesssim0.1$, 
the
flow is well coupled
and multicomponent effects are negligible. For larger velocity differences,
$x_\text{ip}\gtrsim0.1$, frictional heating becomes important 
%
(Fig.~\ref{55-01Z034k}), and for $x_\text{ip}\gtrsim1$ 
the
wind components may
decouple. After decoupling, hydrogen and helium leave the star if the decoupling
occurs at 
%
velocities larger than the escape speed, or fall back on the
stellar surface if the decoupling occurs at 
%
velocities lower than the escape
speed \citep[e.g.][]{ufo,viktor}. 

\begin{figure}[tp]
\includegraphics[width=0.45\textwidth]{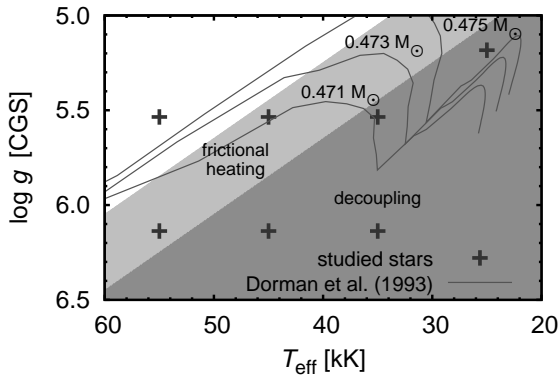}
\caption{%
Multicomponent 
effects
in the $T_\text{eff}-\log g$ diagram}
\label{sitpotvic}
\end{figure}

The parameter domains,
in which multicomponent effects are important are
given in Fig.~\ref{sitpotvic}. From our calculations we conclude \citep[in
agreement with][]{unglaub} that coupled subdwarf winds with
\linebreak
$\dot
M\lesssim10^{-12}\,\text{M}_\odot\,\text{yr}^{-1}$ do not exist.

\section{Inefficient shock cooling}

\begin{figure}[t]
\includegraphics[width=0.45\textwidth]{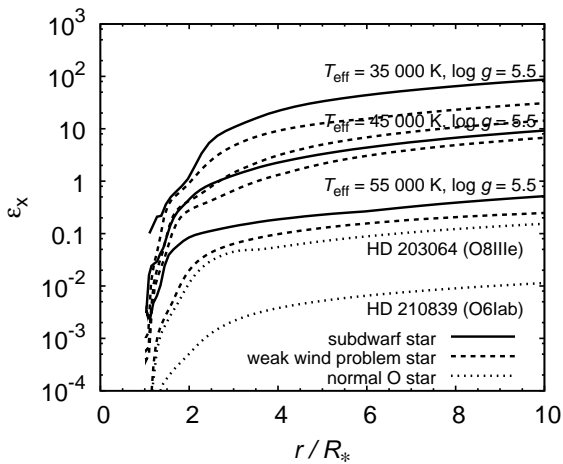}
\caption{%
Dependence of the ratio of cooling and hydrodynamical time scales
on radius for individual stars}
\label{epsix}
\end{figure}

The last similarity between the stellar 
winds
of B stars and subdwarf stars
mentioned here is connected with inefficient shock cooling. Hot star winds are
known to be subject of the instability of radiative driving. As a result of
this, 
a
small part of the wind material of luminous O stars is very hot and emits
X-rays \citep{ocr,felpulpal}.

However, for low-luminosity stars (e.g., the subdwarf stars) 
the
hot part of the
stellar wind may be significantly larger \citep{cobecru,nlteiii}. This can be
seen from 
%
the ratio of cooling and hydrodynamical time scales 
plotted
in
Fig.~\ref{epsix}. For 
%
O stars 
of normal luminosity
this ratio is very small,
indicating that only a very small part of the wind 
%
is very hot.
On the other hand, for stars with weaker winds this ratio could be relatively
large, indicating that once a shock occurs in the wind, the material has too low
density to 
cool down significantly
radiatively. Consequently, for low-luminosity
stars a significant portion of the 
wind
can be very hot, and it
can be difficult to observe such winds.

For low-luminosity O stars this can 
explain
the so-called ``weak
wind problem'' \citep[see][]{nlteiii}. The same effect could be 
also present
in subdwarf stars 
(Fig.~\ref{epsix}).

\section{Similarity to the first stars}

Stellar winds of hot subdwarf stars are also similar to the stellar wind of
the
first stars in the Universe. First stars formed from the pristine gas 
that emerged from
the primordial nucleosynthesis. Consequently, these stars were pure
hydrogen-helium, and were very massive \citep[$M\sim100\,\text{M}_\odot$,
e.g.,][]{ompal} and hot. Hydrogen and helium lines are not able to accelerate
the wind from these stars \citep{cnovit}, however, during the core He-burning
phase the CNO elements (primary nitrogen) emerge 
at
the surface due to mixing
\citep{mee}. Consequently, similarly to subdwarfs stars, evolved first stars had
CNO-driven winds \citep{graham,cnovit}.

\section{Importance of the stellar winds}

Although stellar winds of hot subdwarf stars are rather weak, they can still
have some observable effects.

Weak stellar winds were invoked for the explanation why some subdwarf stars
pulsate whereas others 
do
not \citep[e.g.,][]{fontana}. In agreement with
\citet{unglaub}, our results do not support this explanation, because weak
stellar winds are expected to be decoupled.

On the other hand, weak stellar winds in binaries consisting of a subdwarf star
and a compact object (neutron star or a black hole) may lead to production of
X-rays. Last but not least, interaction of 
the
stellar wind 
from
a subdwarf star with
a planetary companion may also cause observable effects \citep{geier}.

\section{Conclusions}

We have shown that stellar winds 
from
central stars of planetary nebulae resemble
stellar winds of luminous O stars. 
In contrast,
stellar winds of hot
subdwarf stars are much weaker and resemble stellar winds of B~stars in several
important aspects.
They
\begin{itemize}
\item 
%
are driven mainly by lighter elements,
\item 
%
show multicomponent effects,
\item 
%
may be very hot in outer parts due to
inefficient shock cooling.
\end{itemize}

\acknowledgments
This work was supported by grant GA \v{C}R
205/07/0031. The Astronomical Institute Ond\v{r}ejov is supported by the project
AV0\,Z10030501.

\nocite{*}
\bibliographystyle{spr-mp-nameyear-cnd}
\bibliography{biblio-u1}

\end{document}